\documentclass[a4paper]{llncs}
\pdfoutput=1
\usepackage[utf8]{inputenc}
\usepackage[english]{babel}
\usepackage{microtype}
\usepackage{booktabs}
\usepackage{graphicx}
\usepackage{setspace}
\usepackage{cite} 
\usepackage{listings}
\usepackage[hidelinks]{hyperref}
\usepackage[capitalise]{cleveref}
\crefname{line}{line}{lines} 

\begin{document}
\pagestyle{headings}  
\mainmatter              
\title{Generating events with style}
\author{Matthieu Boutier\inst{1} \and Gabriel Kerneis\inst{1,2}}

\newcommand{\lcb}{{\tt {\char '173}}}  
\newcommand{\rcb}{{\tt {\char '175}}}  
\institute{PPS, Université Paris Diderot, 
      \email{\lcb first.last\rcb @pps.univ-paris-diderot.fr}
  \and
  University of Cambridge}

\maketitle              

\begin{abstract}
Threads and events are two common abstractions for writing concurrent programs.
Because threads are often more convenient, but events more efficient, it is
natural to want to translate the former into the latter.  However, whereas there
are many different event-driven styles, existing translators often apply ad-hoc
rules which do not reflect this diversity.

We analyse various control-flow and data-flow encodings in real-world
event-driven code, and we observe that it is possible to generate any of these
styles automatically from threaded code, by applying certain carefully chosen
classical program transformations.  In particular, we implement two of these
transformations, lambda lifting and environments, in CPC, an extension of the C
language for writing concurrent systems.  Finally, we find out that, although
rarely used in real-world programs because it is tedious to perform manually,
lambda lifting yields better performance than environments in most of our
benchmarks.

\keywords{Concurrency, program transformations, event-driven style}
\end{abstract}

\section{Introduction}

Most computer programs are \emph{concurrent} programs, which need to perform
several tasks at the same time.  For example, a network server needs to serve
multiple clients at a time; a GUI needs to handle multiple keyboard and mouse
inputs; and a network program with a graphical interface (e.g.\ a web browser)
needs to do both simultaneously.

\subsubsection{Translating threads into events}

There are many different techniques to implement concurrent programs.  A very
common abstraction is provided by \emph{threads}, or \emph{lightweight
processes}.  In a threa\-ded program, concurrent tasks are executed by a number of
independent threads which communicate through a shared memory heap.  An
alternative to threads is \emph{event-driven} programming.  An event-driven
program interacts with its environment by reacting to a set of stimuli called
\emph{events}.  At any given point in time, to every event is associated a piece
of code known as the \emph{handler} for this event.  A global scheduler, known
as the \emph{event loop}, repeatedly waits for an event to occur and invokes the
associated handler.  Performing a complex task requires to coordinate several
event handlers by exchanging appropriate events.

Unlike threads, event handlers do not have an associated stack; event-driven
programs are therefore more lightweight and often faster than their threaded
counterparts.  However, because it splits the flow of control into multiple tiny
event handlers, event-driven programming is generally deemed more difficult and
error-prone.  Additionally, event-driven programming alone is often not powerful
enough, in particular when accessing blocking APIs or using multiple processor
cores; it is then necessary to write \emph{hybrid} code, that uses both
\emph{preemptively-scheduled} threads and \emph{cooperatively-scheduled} event
handlers, which is even more difficult.

Since event-driven programming is more difficult but more efficient than
threaded programming, it is natural to want to at least partially automate it.
\emph{Continuation-Passing C} (CPC \cite{kerneis2011}) is an extension of the C
programming language for writing concurrent systems.  The CPC programmer
manipulates very lightweight threads, choosing whether they should be
cooperatively or preemptively scheduled at any given point.  The CPC program is
then processed by the \emph{CPC translator}, which produces highly efficient
sequentialised event-loop code, and uses native threads to execute the
preemptive parts.  The translation from threads into events is performed by a
series of classical source-to-source program transformations: splitting of the
control flow into mutually recursive inner functions, lambda lifting of these
functions created by the splitting pass, and CPS conversion of the resulting
code.  This approach retains the best of both worlds: the relative convenience
of programming with threads, and the low memory usage of event-loop code.

\subsubsection{The many styles of events}

Not all event-driven programs look the same: several styles and implementations
exist, depending on the programmer's taste.  Since event-driven programming
consists in manually handling the control flow and data flow of each task, a
tedious and error-prone activity, the programmer often choses a style based on
some trade-off between (his intuition of) efficiency and code-readability,
and then sticks with it in the whole program.  Even if the representation of
control or data turns out to be suboptimal, changing it would generally require
a complete refactoring of the program, not likely to be undertaken for an
uncertain performance gain.  In large event-driven programs, written by several
people or over a long timespan, it is even possible to find a mix of several
styles making the code even harder to decipher.

For example, the transformations performed by the CPC translator yield
event-driven code where control flow is encoded as long, intricate chains of
callbacks, and where local state is stored in tiny data structures, repeatedly
copied from one event-handler to the next.  We can afford these techniques
because we generate the code automatically. Hand-written programs often use less
tedious approaches, such as state machines to encode control flow and coarse
long-lived data structures to store local state; these are easier to
understand and debug but might be less efficient.  Since the transformations
performed by the CPC translator are completely automated, it offers an ideal
opportunity to generate several event-driven variants of the same threaded
program, and compare their efficiency.

\subsubsection{Contributions}

We first review existing translators from threads to events
(\cref{sec:related}), and analyse several examples of event-driven styles found
in real-world programs (\cref{sec:flow}).  We identify two typical kinds of
control-flow and data-flow encodings: callbacks or state machines for the
control flow, and coarse-grained or minimal data structures for the data flow.

We then propose a set of automatic program transformations to produce each of
these four variants (\cref{sec:generating}).  Control flow is translated by
splitting and CPS conversion to produce callbacks; adding a pass of
defunctionalisation yields state machines.  Data flow is translated either by
lambda lifting, to produce minimal, short-lived data structures, or using shared
environments for coarse-grained ones.

Finally, we implement eCPC, a variant of CPC using shared environments instead of
lambda lifting to handle the data flow in the generated programs
(\cref{sec:ecpc}).  We find out that, although rarely used in real-world
event-driven programs because it is tedious to perform manually, lambda lifting
yields faster code than environments in most of our benchmarks.  To the best of
our knowledge, CPC is currently the only threads-to-events translator using
lambda lifting.

\section{Related work}
\label{sec:related}

The translation of threads into events has been rediscovered many times
\cite{DBLP:conf/sensys/DunkelsSVA06,DBLP:conf/usenix/KrohnKK07,weave}.  In this
section, we review existing solutions, and observe that each of them generates
only one particular kind of event-driven style.  As we shall see in
\cref{sec:generating}, we believe that these implementations are in fact a few
classical transformation techniques, studied extensively in the context of
functional languages, and adapted to imperative languages, sometimes
unknowingly, by programmers trying to solve the issue of writing events in a
threaded style.

The first example known to us is \emph{Weave}, an unpublished tool used at IBM
in the late 1990's to write firmware and drivers for SSA-SCSI RAID
storage adapters \cite{weave}.  It translates annotated Woven-C code, written in
threaded style, into C code hooked into the underlying event-driven kernel.

Adya et al.\ \cite{DBLP:conf/usenix/AdyaHTBD02} provide a detailed analysis of
control flow in threads and events programs, and implement adaptors between
event-driven and threaded code to write hybrid programs mixing both styles.

Duff introduces a technique, known as \emph{Duff's device} \cite{duff}, to
express general loop unrolling directly in C, using the \texttt{switch}
statement.  Much later, this technique has been employed multiple times to
express state machines and event-driven programs in a threaded style:
\emph{protothreads} \cite{DBLP:conf/sensys/DunkelsSVA06}, \emph{FairThreads}'
automata \cite{DBLP:journals/concurrency/Boussinot06}.  These libraries help
keep a clearer flow of control but they provide no automatic handling of data
flow: the programmer is expected to save local variables manually in his own
data structures, just like in event-driven style.

\emph{Tame} \cite{DBLP:conf/usenix/KrohnKK07} is a C++ language extension and
library which exposes events to the programmer but does not impose event-driven
style: it generates state machines to avoid the stack ripping issue and retain a
thread-like feeling.  Similarly to Weave, the programmer needs to annotate local
variables that must be saved across context switches.

\emph{TaskJava} \cite{DBLP:conf/pepm/FischerMM07} implements the same idea as
Tame, in Java, but preserves local variables automatically, storing them in a
state record.  \emph{Kilim} \cite{DBLP:conf/ecoop/SrinivasanM08} is a
message-passing framework for Java providing actor-based, lightweight threads.
It is also implemented by a partial CPS conversion performed on annotated
functions, but contrary to TaskJava, it works at the JVM bytecode level.

\emph{MapJAX} \cite{DBLP:conf/usenix/MyersCCL07} is a conservative extension of
Javascript for writing asynchronous RPC, compiled to plain Javascript using some
kind of ad-hoc splitting and CPS conversion.  Interestingly enough, the authors
note that, in spite of Javascript's support for nested functions, they need to
perform ``function denesting'' for performance reasons; they store free
variables in environments (``closure objects'') rather than using
lambda lifting.

\emph{AC} \cite{DBLP:conf/oopsla/HarrisAIM11} is a set of language constructs
for composable asynchronous I/O in C and C++.  Harris et al.\ introduce
\texttt{do..finish} and \texttt{async} operators to write asynchronous requests
in a synchronous style, and give an operational semantics.  The language
constructs are somewhat similar to those of Tame but the implementation is very
different, using LLVM code blocks or macros based on GCC's nested functions
rather than source-to-source transformations.

\section{Control flow and data flow in event-driven code}
\label{sec:flow}

Because event-driven programs do not use the native call stack to store return
addresses and local variables, they must encode the control flow and data flow
in data structures, the bookkeeping of which is the programmer's responsibility.
This yields a diversity of styles among event-driven programs, depending on the
programmer's taste, creativity, and his perception of efficiency.  In this
section, we analyse how control flow and data flow are encoded in several
examples of real-world event-driven programs, and compare them to equivalent
threaded-style programs.

\subsection{Control flow}
\label{sec:control-flow}

Two main techniques are used to represent the control flow in event-driven
programming: callbacks and state machines.

\paragraph{Callbacks}

Most of the time, control flow is implemented with \emph{callbacks}.  Instead
of performing a blocking function call, the programmer calls a non-blocking
equivalent that cooperates with the event loop, providing a function pointer
to be called back once the non-blocking call is done.  This callback function
is actually the continuation of the blocking operation.

Developing large programs raises the issue of composing event handlers.
Whereas threaded code has return addresses stored on the stack and a standard
calling sequence to coordinate the caller and the callee,
event-driven code needs to define its own strategy to layer callbacks, storing
the callback to the next layer in some data structure associated with the
event handler.  The ``continuation stack'' of callbacks is often split in
various places of the code, each callback encoding its chunk of the stack in
an ad-hoc manner.

Consider for instance the accept loop of an HTTP server that accepts clients and
starts two tasks for each of them: a client handler, and a timeout to disconnect
idle clients.  With cooperative threads, this would be implemented as a mere
infinite loop with a cooperation point.  The following code is an example of
such an accept loop written with CPC.
\begin{lstlisting}
cps int cpc_accept(int fd) {
  cpc_io_wait(fd, CPC_IO_IN);
  return accept(fd, NULL, NULL);
}
cps int accept_loop(int fd) {
  int client_fd;
  while(1) {
    client_fd = cpc_accept(fd);
    cpc_spawn httpTimeout(client_fd, clientTimeout);
    cpc_spawn httpClientHandler(client_fd);
  }
}
\end{lstlisting}
The programmer calls \texttt{cpc\_spawn accept\_loop(fd)} to create a new thread
that runs the accept loop; the function \texttt{accept\_loop} then waits for
incoming connections with the cooperating primitive \texttt{cpc\_io\_wait}, and
creates two new threads for each client (\texttt{httpTimeout} and
\texttt{httpClientHandler}), which kill each other upon completion.  Note that
cooperative functions are annotated with the \texttt{cps} keyword; such
\emph{cps} functions are to be converted into event-driven style by the CPC
translator.

\Cref{fig:polipo} shows the (very simplified) code of the accepting loop in
Polipo, a caching web-proxy written by
Chroboczek.\footnote{\url{http://www.pps.univ-paris-diderot.fr/~jch/software/polipo/}.}
This code is equivalent to the threaded version above, and uses several levels
of callbacks.

\begin{figure}[tbp]
\begin{spacing}{0.8}
\begin{lstlisting}[numbers=left,xleftmargin=0pt]
FdEventHandlerPtr
schedule_accept(int fd,
    int (*handler)(int, FdEventHandlerPtr, AcceptRequestPtr),
    void *data) {
  FdEventHandlerPtr event;
  AcceptRequestRec request;
  int done;

  request.fd = fd;
  request.handler = handler;                        /*@\label{l:pol1}@*/
  request.data = data;
  event = registerFdEvent(fd, POLLOUT|POLLIN,       /*@\label{l:pol2}@*/
              do_scheduled_accept,
              sizeof(request), &request);
  return event;
}

int
do_scheduled_accept(int status, FdEventHandlerPtr event) {
  AcceptRequestPtr request = (AcceptRequestPtr)&event->data;
  int rc, done;

  rc = accept(request->fd, NULL, NULL);             /*@\label{l:pol3}@*/
  done = request->handler(rc, event, request);      /*@\label{l:pol4}@*/
  return done;
}

int
httpAccept(int fd, FdEventHandlerPtr event,
           AcceptRequestPtr request) {
  HTTPConnectionPtr connection;
  TimeEventHandlerPtr timeout;

  connection = httpMakeConnection();/*@\label{l:pol7}@*/
  timeout = scheduleTimeEvent(clientTimeout,        /*@\label{l:pol5}@*/
                httpTimeoutHandler,
                sizeof(connection), &connection);
  connection->fd = fd;
  connection->timeout = timeout;
  connection->flags = CONN_READER;
  do_stream_buf(IO_READ | IO_NOTNOW,                /*@\label{l:pol6}@*/
         connection->fd, 0, &connection->reqbuf,
          CHUNK_SIZE, httpClientHandler, connection);
  return 0;
}
\end{lstlisting}
\end{spacing}
\caption{Accept loop callbacks in Polipo (simplified)}
\label{fig:polipo}
\end{figure}

In Polipo, the accept loop is started by a call to
\texttt{schedule\_\-accept(fd, http\-Accept, NULL)}.
This function stores the pointer to the (second-level) callback
\texttt{httpAccept} in the \texttt{handler} field of the \texttt{request} data
structure (\cref{l:pol1}), and registers a (first-level) callback to
\texttt{do\_scheduled\_accept}, through
\texttt{regi\-ster\-Fd\-Event}.  Each time the file descriptor \texttt{fd}
becomes ready (not shown), the event loop calls the (first-level) callback
\texttt{do\_scheduled\_accept}, which performs the actual \texttt{accept} system
call (\cref{l:pol3}) and finally invokes the (second-level) callback \texttt{httpAccept}
stored in \texttt{request->handler} (\cref{l:pol4}).

This callback schedules two new event handlers, \texttt{httpTimeout} and
\texttt{http\-Client\-Handler}.  The former is a timeout handler, registered by
\texttt{schedule\-Time\-Event} (\cref{l:pol5}); the latter reacts I/O events to read requests
from the client, and is registered by \texttt{do\_stream\_buf} (\cref{l:pol6}).  Note that
those helper functions that register callbacks with the event loop use other
intermediary callbacks themselves, just like \texttt{schedule\_accept} uses
\texttt{do\_schedule\_accept}.

In the original Polipo code, things are even more complex since
\texttt{schedule\_\allowbreak accept} is called from \texttt{httpAcceptAgain}, yet another
callback that is registered by \texttt{httpAccept} itself in some error cases.
The control flow becomes very hard to follow, in particular when errors are
triggered: each callback must be prepared to cope with error codes, or to
follow-up the unexpected value to the next layer.  In some parts of the code,
this style looks a lot like an error monad manually interleaved with a
continuation monad.  Without a strict discipline and well-defined conventions
about composition, the flexibility of callbacks easily traps the programmer in a
control-flow and storage-allocation maze.

\paragraph{State machines}

When the multiplication of callbacks becomes unbearable, the
event-loop programmer might refactor his code to use a state machine.
Instead of splitting a computation into as many callbacks as it has atomic
steps, the programmer registers a single callback that will be called over and
over until the computation is done.  This callback implements a state machine:
it stores the current state of the computation into an ad-hoc data structure,
just like threaded code would store the program counter, and uses it upon
resuming to jump to the appropriate location.

\Cref{fig:transmission} shows how the initial handshake of a BitTorrent
connection is handled in
\emph{Transmission},\footnote{\url{http://www.transmissionbt.com/}.} a popular
and efficient BitTorrent client written in (mostly) event-driven style.  Until
the handshake is over, all data arriving from a peer is handed over by the
event loop to the \texttt{canRead} callback.  This function implements a state
machine, whose state is stored in the \texttt{state} field of a \texttt{handshake}
data structure.  This field is initialised to \texttt{AWAITING\_HANDSHAKE} when
the connection is established (not shown) and updated by the functions
responsible for each step of the handshake.

\begin{figure}[tbp]
\begin{spacing}{0.8}
\begin{lstlisting}[numbers=left,xleftmargin=0pt]
static ReadState
canRead(struct evbuffer *inbuf, tr_handshake *handshake) {
  ReadState ret = READ_NOW;

  while(ret == READ_NOW) {
    switch(handshake->state) {
      case AWAITING_HANDSHAKE:                      /*@\label{l:tr1}@*/
        ret = readHandshake  (handshake, inbuf);
        break;
      case AWAITING_PEER_ID:                        /*@\label{l:tr7}@*/
        ret = readPeerId    (handshake, inbuf);
        break;
      /* ... cases dealing with encryption omitted */
    }
  }
  return ret;                                       /*@\label{l:tr3}@*/
}

static int
readHandshake(tr_handshake *handshake,
              struct evbuffer *inbuf) {
  uint8_t pstr[20], reserved[HANDSHAKE_FLAGS_LEN],
      hash[SHA_DIGEST_LENGTH];

  if(evbuffer_get_length(inbuf) < INCOMING_HANDSHAKE_LEN)
    return READ_LATER;                              /*@\label{l:tr2}@*/
  tr_peerIoReadBytes(handshake->io, inbuf, pstr, 20);
  if(memcmp(pstr, "\023BitTorrent protocol", 20))   /*@\label{l:tr4}@*/
    return tr_handshakeDone(handshake, false);
  tr_peerIoReadBytes(handshake->io, inbuf, reserved, ...);
  tr_peerIoReadBytes(handshake->io, inbuf, hash, ...);
  /* ... parsing of handshake and sending reply omitted  */
  handshake->state = AWAITING_PEER_ID;              /*@\label{l:tr5}@*/
  return READ_NOW;                                  /*@\label{l:tr6}@*/
}

static int
readPeerId(tr_handshake *handshake, struct evbuffer *inbuf) {
  uint8_t peer_id[PEER_ID_LEN];

  if(evbuffer_get_length(inbuf) < PEER_ID_LEN)      /*@\label{l:tr8}@*/
    return READ_LATER;
  tr_peerIoReadBytes(handshake->io, inbuf, peer_id, ...);
  /* ... parsing of peer id omitted */
  return tr_handshakeDone(handshake, true);         /*@\label{l:tr9}@*/
}
\end{lstlisting}
\end{spacing}
\caption{Handshake state-machine in Transmission (simplified)}
\label{fig:transmission}
\end{figure}

The first part of the handshake is dispatched by \texttt{canRead} to the
\texttt{read\-Hand\-shake} function (\cref{l:tr1}).  It receives the buffer \texttt{inbuf}
containing the bytes received so far; if not enough data has yet been received
to carry on the handshake, it returns \texttt{READ\_LATER} to \texttt{canRead} (\cref{l:tr2}),
which forwards it to the event loop to be called back when more data is
available (\cref{l:tr3}).  Otherwise, it checks the BitTorrent header (\cref{l:tr4}), parses the
first part of the handshake, registers a callback to send a reply handshake
(not shown), and finally updates the state (\cref{l:tr5}) and returns \texttt{READ\_NOW} to
indicate that the rest of the handshake should be processed immediately (\cref{l:tr6}).

Note what happens when the BitTorrent header is wrong (\cref{l:tr4}): the function
\texttt{tr\_hand\-shake\-Done} is called with \texttt{false} as its second parameter,
indicating that some error occurred.  This function (not shown) is responsible
for invoking the callback \texttt{hand\-shake\-->\-doneCB} and then deallocating the
\texttt{handshake} structure.  This is another example of the multiple layers of
callbacks mentioned above.

If the first part of the handshake completes without error, \texttt{canRead}
then dispatches the buffer to \texttt{readPeerId} which completes the handshake
(\cref{l:tr7}).  Just like \texttt{readHandshake}, it returns \texttt{READ\_LATER} if the
second part of the handshake has not arrived yet (\cref{l:tr8}) and finally calls
\texttt{tr\_handshakeDone} with \texttt{true} to indicate that the handshake has
been successfully completed (\cref{l:tr9}).

In the original code, ten additional states are used to deal with the various
steps of negotiating encryption keys.  The last of these steps finally rolls
back the state to \texttt{AWAITING\_HANDSHAKE} and the keys are used by the
function \texttt{tr\_peerIoReadBytes} to decrypt the rest of the exchange
transparently.  The state machine approach makes the code slightly more
readable than using pure callbacks.

\subsection{Data flow}
\label{sec:data-flow}

Since each callback function performs only a small part of the whole
computation, the event-loop programmer needs to store temporary data
required to carry on the computation in heap-allocated data structures,
whereas stack-allocated variables would sometimes seem more natural in
threaded style.  The content of these data structures depends heavily on the
program being developed but we can characterise some common patterns.

Event loops generally provide some means to specify a \texttt{void*} pointer
when registering an event handler.  When the expected event triggers, the
pointer is passed as a parameter to the callback function, along with
information about the event itself.  This allows the programmer to store
partial results in a structure of his choice, and recover it through the
pointer without bothering to maintain the association between event handlers
and data himself.

\paragraph{Coarse-grained, long lived data structures}

These data structures are usually large and coarse-grained.  Each of them
correponds to
some meaningful object in the context of the program, and is passed from
callback to callback through a pointer.  For instance, the \texttt{connection}
structure used in Polipo (\Cref{fig:polipo}) is allocated by
\texttt{httpMakeConnection} when a connection starts (\cref{l:pol7}) and passed
to the callbacks \texttt{httpTimeoutHandler} and \texttt{httpClientHandler}
through the registering functions \texttt{scheduleTimeEvent} (\cref{l:pol5}) and
\texttt{do\_stream\_buf} (\cref{l:pol6}). It lives as long as the HTTP
connection it describes and contains no less than 22 fields: \texttt{fd},
\texttt{timeout}, \texttt{buf}, \texttt{pipelined},~\textit{etc}. The
\texttt{tr\_handshake} structure passed to \texttt{canRead} in Transmission is
similarly large, with 18 fields.

Some of these fields need to live for the whole connection (eg.\ \texttt{fd}
which stores the file descriptor of the socket) but others are used only
transiently (eg.\ \texttt{buf} which is filled only when sending a reply), or
even not at all in some cases (eg.\ the structure \texttt{HTTPConnectionPtr} is
used for both client and server connections, but the \texttt{pipelined} field is
never used in the client case).  Even if it wastes memory in some cases, it
would be too much of a hassle for the programmer to track every possible data
flow in the program and create ad-hoc data structures for each of them.

\paragraph{Minimal, short-lived data structures}

In some simple cases, however, the event-loop programmer is able to allocate
very small and short-lived data structures.  These minimal data structures are
allocated directly within an event handler and are deallocated when the
associated callback returns.  They might even be allocated on the stack by the
programmer and copied inside the event-loop internals by the helper function
registering the event handler.  The overhead is therefore kept as low as
possible.

For instance, the function \texttt{schedule\_accept} passes a tiny,
stack-allocated structure \texttt{request} to the helper function
\texttt{registerFdEvent} (\cref{fig:polipo}, \cref{l:pol2}).  This structure is
of type \texttt{AcceptRequestRec} (not shown), which contains only three fields:
an integer \texttt{fd} and two pointers \texttt{handler} and \texttt{data}.  It
is copied by \texttt{registerFdEvent} in the event-loop data structure
associated with the event, and freed automatically after the callback
\texttt{do\_\allowbreak scheduled\_\allowbreak accept} has returned; it is as
short-lived and (almost) as compact as possible.

As it turns out, creating truly minimal structures is hard:
\texttt{AcceptRequestRec} could in fact be optimised to get rid off the fields
\texttt{data}, which is always \texttt{NULL} in practice in Polipo, and
\texttt{fd}, which is also present in the encapsulating \texttt{event} data
structure.  Finding every such redundancy in the data flow of a large
event-driven program would be a daunting task, hence the spurious and
redundant fields used to lighten the programmer's burden.

\section{Generating various event-driven styles}
\label{sec:generating}

In this section, we first demonstrate the effect of CPC transformation passes on
a small example; we show that code produced by the CPC translator is very close
to event-driven code using callbacks for control flow, and minimal data
structures for data flow (\cref{sec:cpc}).  We then show how two other classical
translation passes produce different event-driven styles:
\emph{defunctionalising} inner function yields state machines
(\cref{sec:defun}), and encapsulating local variables in \emph{shared
environments} yields larger, long-lived data structures with full context
(\cref{sec:env}).

\subsection{The CPC compilation technique}
\label{sec:cpc}

Consider the following function, which counts seconds down from an initial
value~\texttt{x} to zero.
\begin{lstlisting}
cps void countdown(int x) {
  while(x > 0) {
    printf("%d\n", x--);
    cpc_sleep(1);
  }
  printf("time is over!\n");
}
\end{lstlisting}
This function is annotated with the \texttt{cps} keyword to indicate that it
yields to the CPC scheduler.  This is necessary because it calls the CPC
primitive \texttt{cpc\_sleep}, which also yields to the scheduler.

The CPC translator is structured in a series of proven source-to-source
transformations \cite{kerneis2011}, which turn a threaded-style CPC program into
an equivalent event-driven C program.  \emph{Boxing} first encapsulates a small
number of variables in environments.  \emph{Splitting} then splits the flow of
control of each cps function into a set of inner functions.
\emph{Lambda lifting} removes free local variables introduced by the splitting
step; it copies them from one inner function to the next, yielding closed inner
functions.  Finally, the program is in a form simple enough to perform a
one-pass partial \emph{CPS conversion}.  The resulting continuations are used at
runtime to schedule threads.

In the rest of this section, we show how splitting, lambda lifting and CPS
conversion transform the function \texttt{countdown}.  The boxing pass has no
effect on this example because it only applies to \emph{extruded} variables, the
address of which is retained by the ``address of'' operator (\verb+&+).

\paragraph{Splitting}

The first transformation performed by the CPC translator is \emph{splitting}.
Splitting has been first described by van Wijngaarden for Algol~60
\cite{wijngaarden}, and later adapted by Thielecke to C, albeit in a restrictive
way \cite{DBLP:journals/sigact/Thielecke99}.  It translates control structures 
into mutually recursive functions.

Splitting is done in two steps.  The first step consists in replacing every
control-flow structure, such as \texttt{for} and \texttt{while} loops, by its
equivalent in terms of \texttt{if} and \texttt{goto}.
\begin{lstlisting}
cps void countdown(int x) {
 loop:
  if(x <= 0) goto timeout;
  printf("%d\n", x--);
  cpc_sleep(1);
  goto loop;
 timeout:
  printf("time is over!\n");
}
\end{lstlisting}
The second step uses the fact that \texttt{goto} are equivalent to tail calls
\cite{Ste76}.  It translates every labelled block into an inner function, and
every jump to that label into a tail call (followed by a \texttt{return}) to
that function.
\begin{figure}[htbp]
\begin{lstlisting}[numbers=left]
cps void countdown(int x) {
 cps void loop() {
   if(x <= 0) { timeout(); return; } /*@\label{l:tp1}@*/
   printf("%d\n", x--);
   cpc_sleep(1); loop(); return; /*@\label{l:tp2}@*/
 }
 cps void timeout() { printf("time is over!\n"); return; } /*@\label{l:tp3}@*/
 loop(); return;
}
\end{lstlisting}
\caption{CPC code after splitting}
\label{lst:split}
\end{figure}

Splitting yields a program where each cps function is split in several mutually
recursive, atomic functions, very similar to event handlers.  Additionally, the
tail positions of these inner functions are always either:
\begin{itemize}
    \item a \texttt{return} statement (for instance, on \cref{l:tp3} in the
        previous example),
    \item a tail call to another cps function (\cref{l:tp1}),
  \item a call to an external cps function followed by a call to an
      inner cps function (\cref{l:tp2}).
\end{itemize}
We recognise the typical patterns of an event-driven program that we studied in
\cref{sec:flow}: respectively returning a value to the upper layer
(\cref{fig:polipo} (4)), calling a function to carry on the current computation
(\cref{fig:transmission} (1)), or calling a function with a callback to resume
the computation once it has returned (\cref{fig:polipo} (2)).

Another effect of splitting is the introduction of free variables, which are
bound to the original encapsulating function rather than the new inner ones.
For instance, the variable \texttt{x} is free in the function \texttt{loop}
above.  Because inner functions and free variables are not allowed in C, we
perform a pass of lambda lifting to eliminate them.

\paragraph{Lambda lifting}

The CPC translator then makes the data flow explicit with a lambda-lifting pass.
Lambda lifting, also called closure conversion, is a standard technique to
remove free variables introduced by Johnsson \cite{DBLP:conf/fpca/Johnsson85}.
It is also performed in two steps: parameter lifting and block floating.

Parameter lifting binds every free variable to the inner function where it
appears (for instance \texttt{x} to \texttt{loop} on \cref{l:ll1} below).  The
variable is also added as a parameter at every call point of the function
(\cref{l:ll2,l:ll3}).
\begin{lstlisting}[numbers=left]
cps void countdown(int x) {
 cps void loop(int x) { /*@\label{l:ll1}@*/
   if(x <= 0) { timeout(); return; }
   printf("%d\n", x--);
   cpc_sleep(1); loop(x); return; /*@\label{l:ll2}@*/
 }
 cps void timeout() { printf("time is over!\n"); return; }
 loop(x); return; /*@\label{l:ll3}@*/
}
\end{lstlisting}
Note that because C is a call-by-value language, lifted parameters are
duplicated rather than shared and this step is not correct in general.
It is however sound in the case of CPC because lifted functions are called in
tail position: they never return, which guarantees that at most one copy of each
parameter is reachable at any given time \cite{kerneis2011}.
Block floating is then a trivial extraction of closed, inner functions at
top-level.

Lambda lifting yields a program where the data is copied from function to
function, each copy living as long as the associated handler.  If some piece of
data is no longer needed during the computation, it will not be copied in the
subsequent handlers; for instance, the variable \texttt{x} is not passed to the
function \texttt{timeout}. Hence, lambda lifting produces short-lived, almost
minimal data structures.

\paragraph{CPS conversion}

Finally, the control flow is made explicit with a CPS conversion
\cite{DBLP:journals/tcs/Plotkin75,DBLP:journals/lisp/Reynolds93}.  The
continuations store callbacks and their parameters in a regular stack-like
structure \texttt{cont} with two primitive operations: \texttt{push} to add a
function on the continuation, and \texttt{invoke} to call the first function of
the continuation.
\begin{lstlisting}
cps void loop(int x, cont *k) {
  if(x <= 0) { timeout(k); return; }
  printf("%d\n", x--);
  cpc_sleep(1, push(loop, x, k)); return;
}
cps void timeout(cont *k) {
  printf("time is over!\n");
  invoke(k); return;
}
cps void countdown(int x, cont *k) { loop(x, k); return; }
\end{lstlisting}
CPS conversion turns out to be an efficient and systematic implementation of the
layered callback scheme described in \cref{sec:control-flow}.
Note that, just like lambda lifting, CPS conversion is not correct in general in
an imperative call-by-value language, because of duplicated variables on the
continuation.  It is however correct in the case of CPC, for reasons similar to
the correctness of lambda lifting \cite{kerneis2011}.

\subsection{Defunctionalising inner functions}
\label{sec:defun}

Defunctionalisation is a compilation technique introduced by Reynolds to
translate higher-order programs into first-order ones \cite{reynolds72}.  It
maps every first-class function to a first-order structure that contains
both an index representing the function, and the values of its free variables.
These data structures are usually a constructor, whose parameters store the free
variables.  Function calls are then performed by a dedicated function that
dispatches on the constructor, restores the content of the free variables and
executes the code of the relevant function.

The dispatch function introduced by defunctionalisation is very close to a state
automaton. It is therefore not surprising that defunctionalising inner functions
in CPC yields an event-driven style similar to state machines
(\cref{sec:control-flow}).

\paragraph{Defunctionalisation of CPC programs}

Usually, defunctionalisation contains an implicit lambda-lifting pass, to make
free variables explicit and store them in constructors.  For example, a function
\texttt{fn x => x + y} would be replaced by an instance of \texttt{LAMBDA of int},
with the free variable \texttt{y} copied in the constructor \texttt{LAMBDA}.  The
dispatch function would then have a case: \texttt{dispatch (LAMBDA y, x) = x + y}.

In this discussion, we wish to decouple this data-flow transformation from the
translation of the control flow into a state machine.  Therefore, we define the
dispatch function as an inner function which merges the content of the other
inner functions but still contains free variables.  This is possible because the
splitting pass does not create any closure: it introduces inner functions with
free variables, but these are always called directly, not stored as first-class
values whose free variables must be captured.

Consider again our countdown example after the splitting pass
(\cref{lst:split}).  Once defunctionalised, it contains a single inner function
\texttt{dispatch} that dispatches on an enumeration representing the former inner
function \texttt{loop} and \texttt{timeout}.
\begin{lstlisting}[numbers=left]
enum state { LOOP, TIMEOUT };
cps void countdown(int x) {
  cps void dispatch(enum state s) {
    switch(s) {
     case LOOP:
      if(x <= 0) { dispatch(TIMEOUT); return; } /*@\label{l:df1}@*/
      printf("%d\n", x--);
      cpc_sleep(1); dispatch(LOOP); return; /*@\label{l:df2}@*/
     case TIMEOUT:
      printf("time is over!\n"); return;
    }
  }
  dispatch(LOOP); return;
}
\end{lstlisting}
As an optimisation, the recursive call to \texttt{dispatch} on \cref{l:df1} can be
replaced by a \texttt{goto} statement.  However, we cannot replace the call that
follows the cps function \texttt{cpc\_sleep(1)} on \cref{l:df2}, since we will need
to provide \texttt{dispatch} as a callback to \texttt{cpc\_sleep} during CPS
conversion, to avoid blocking.

We must then eliminate free variables and inner functions, with a lambda-lifting
pass.  It is still correct because defunctionnalisation does not break the
required invariants on tail calls.  We finally reach code that is similar in
style to the state-machine shown in \cref{fig:transmission}.
\begin{lstlisting}
cps void dispatch(enum state s, int x) {
  switch(s) {
   case LOOP:
    if(x <= 0) goto timeout_label;
    printf("%d\n", x--);
    cpc_sleep(1); dispatch(LOOP, x); return;
   case TIMEOUT: timeout_label:
    printf("time is over!\n"); return;
  }
}
cps void countdown(int x) { dispatch(LOOP, x); return; }
\end{lstlisting}
In this example, we have also replaced the first occurrence of \texttt{dispatch}
with \texttt{goto timeout\_\-label}, as discussed above, which avoids the final
function call when the counter reaches zero.

If we ignore the \texttt{switch} -- which serves mainly as an entry point to the
dispatch function, à la Duff's device \cite{duff} -- we recognise the
intermediate code generated during the first step of splitting, as having an
explicit control flow using gotos but without inner functions.  In retrospect,
the second step of splitting, which translates gotos to inner functions, can be
considered as a form a \emph{refunctionalisation}, the left-inverse of
defunctionalisation \cite{DBLP:conf/ppdp/DanvyN01}.

\paragraph{Benefits}

The translation presented here is in fact a \emph{partial} defunctionalisation:
each cps function in the original program gets its own dispatch function, and
only inner functions are defunctionalised.  A global defunctionalisation would
imply a whole program analysis, would break modular compilation, and would probably
not be very efficient because C compilers are optimised to compile hand-written,
reasonably-sized functions rather than a giant state automaton with hundreds of
states.  On the other hand, since it is only partial, this translation does not
eliminate the need for a subsequent CPS conversion step to translate calls to
external cps functions into operations on continuations.

Despite adding a new translation step while keeping the final CPS conversion,
this approach has several advantages over the CPS conversion of many smaller,
mutually recursive functions performed by the current CPC translator.  First,
we do not pay the cost of a CPS call for inner functions.  This might bring
significant speed-ups in the case of tight loops or complex control flows.
Moreover, it leaves with much more optimisation opportunities for the C
compiler, for instance to store certain variables in registers, and reduces
the number of operations on the continuations.  It also makes debugging
easier, avoiding numerous hops through ancillary cps functions.

\subsection{Shared environments}
\label{sec:env}

The two main compilation techniques to handle free variables are
lambda lifting, illustrated in \cref{sec:cpc} and discussed extensively in a
previous article \cite{kerneis2011}, and \emph{environments}.  An environment is
a data structure used to capture every free variable of a first-class function
when it is defined; when the function is later applied, it accesses its
variables through the environment.  Environments add a layer of indirection, but
contrary to lambda lifting they do not require free variables to be copied on every
function call.

In most functional languages, each environment represents the free variables of a
single function; a pair of a function pointer and its environment is called a
closure.  However, nothing prevents in principle an environment from being
\emph{shared} between several functions, provided they have the same free
variables.  We use this technique to allocate a single environment shared by
inner functions, containing all local variables and function parameters.

\paragraph{An example of shared environments}

Consider once again our countdown example after splitting (\cref{lst:split}).
We introduce an environment to contain the local variables of \texttt{countdown}
(here, there is only \texttt{x}).
\begin{lstlisting}[numbers=left]
struct env_countdown { int x };
cps void countdown(int x) {
  struct env_countdown *e =
     malloc(sizeof(struct env_countdown)); /*@\label{l:env1}@*/
  e->x = x;  /*@\label{l:env2}@*/
  cps void loop(struct env_countdown *e) {
    if(e->x <= 0) { timeout(); return; } /*@\label{l:env3}@*/
    printf("%d\n", e->x--); /*@\label{l:env4}@*/
    cpc_sleep(1); loop(e); return; /*@\label{l:env5}@*/
  }
  cps void timeout(struct env_countdown *e) {
    printf("time is over!\n");
    free(e); return; /*@\label{l:env6}@*/
  }
  loop(e); return;
}
\end{lstlisting}
The environment is allocated (\cref{l:env1}) and initialised (\cref{l:env2})
when the function \texttt{countdown} is entered.  The inner functions access
\texttt{x} through the environment, either to read (\cref{l:env3}) or to write it
(\cref{l:env4}).  A pointer to the environment is passed from function to
function (\cref{l:env5}); hence all inner functions share the same environment.
Finally, the environment is deallocated just before the last inner function
exits (\cref{l:env6}).

The resulting code is similar in style to hand-written event-driven code, with a
single, heap-allocated data structure sharing the local state between a set of
callbacks.  Note that inner functions have no remaining free variable and can
therefore be lambda-lifted trivially.

\paragraph{Benefits}

Encapsulating local variables in environments avoids having to copy them back
and forth between the continuation and the native call stack.   However, it does
not necessarily mean that the generated programs are faster; in fact,
lambda-lifted programs are often more efficient (\cref{sec:ecpc}).
Another advantage of environments is that they make programs easier to debug, because
the local state is always fully available, whereas in a lambda-lifted program
``useless'' variables are discarded as soon as possible, even though they might
be useful to understand what went wrong before the program crashed.

\section{Evaluation}
\label{sec:ecpc}

In this section, we describe the implementation of eCPC, a CPC variant using
shared environments instead of lambda lifting to encapsulate the local state of
cps functions.  We then compare the efficiency of programs generated with eCPC
and CPC, and show that the latter is more efficient in most cases.  This
demonstrates the benefits of generating events automatically: most real-world
event-driven programs are based on environments, because they are much easier to
use, although systematic lambda lifting would probably yield faster code.

\subsection{Implementation}

The implementation of eCPC is designed to reuse as much of the existing CPC
infrastructure as possible.  The eCPC translator introduces two new passes:
\emph{preparation} and \emph{generation} of environments.   The former replaces
the boxing pass; the latter replaces lambda lifting.

\paragraph{Environment preparation}

Environments must be introduced before the splitting pass for two reasons.
First, it is easier to identify the exit points of cps functions, where the
environments must be deallocated, before they are split into multiple, mutually
recursive, inner functions.  Furthermore, these environment deallocations occur
in tail position, and have therefore an impact on the splitting pass itself.

Although deallocation points are introduced before splitting, neither allocation
nor initialisation or indirect memory accesses are performed at this stage.
Environments introduced during this preparatory pass are empty shells, of type
\texttt{void*}, that merely serve to mark the deallocation points.  This is
necessary because not all temporary variables have been introduced at this
stage (the splitting pass will generate more of them).  Deciding which variables
will be stored in environments is delayed to a later pass.

This preparatory pass also needs to modify how return values are handled.  In
the original CPC, return values are written directly in the continuation when
the returning function invokes its continuation.  This is made possible by the
convention that the return value of a cps function is the last parameter of its
continuation, hence at a fixed position on the continuation stack.  Such is not
the case in eCPC, where function parameters are kept in the environment rather
than copied on the continuation.

In eCPC, the caller function passes a pointer to the callee, indicating the
address where the callee must write its return value.\footnote{Note that a
similar device would be necessary to implement defunctionalisation, because the
\texttt{dispatch} function is a generic callback which might receive many
different types of return values.} The preparatory pass transforms every cps
function returning a type $T$ (different from \texttt{void}) into a function
returning \texttt{void} with an additional parameter of type $T*$; call and return
points are modified accordingly.  The implementation of CPC primitives in the
CPC runtime is also modified to reflect this change.

\paragraph{Environment generation}

After the splitting pass, the eCPC translator allocates and initialises
environments, and replaces variables by their counterpart in the environment.

First, it collects local variables (except the environment pointer itself) and
function parameters and generates the layout of the associated environment.
Then, it allocates the environment and initialises the fields corresponding to
the function parameters.  Because this initialisation is done at the very
beginning of the translated function, it does not affect the tails, thus
preserving the correctness of CPS conversion.  Finally, every use of variables
is replaced by its counterpart in the environment, local variables are
discarded, and inner functions are modified to receive the environment as a
parameter instead.

The CPS conversion is kept unchanged: the issue of return values is dealt with
completely in the preparatory pass and every cps function returns \texttt{void} at
this stage.

\subsection{Benchmark results}
\label{sec:ecpc-bench}

We previously designed a set of benchmarks to compare CPC to other thread
libraries, and have shown that CPC is as fast as the fastest thread libraries
available to us while providing at least an order of magnitude more threads
\cite{kerneis2011}.  We reuse these benchmarks here to compare the speed of CPC
and eCPC; our experimental setup is unchanged, and detailed in our previous
work.

\paragraph{Primitive operations}

We first measure the time of individual CPC primitives.  \Cref{tab:speed} shows
the relative speed of eCPC compared with CPC for each of our micro-benchmarks:
$t_{\mbox{\tiny eCPC}} / t_{\mbox{\tiny CPC}}$.  A value greater than 1 indicates that eCPC
is slower than CPC.  The slowest primitive operation in CPC is a cps function
call (\emph{cps-call}), mostly because of the multiple layers of indirection
introduced by continuations.  This overhead is even larger in the case of eCPC:
performing a cps function call is 2~to 3~times slower than with CPC.
\begin{table}[htb]
\caption{Ratio of speeds of eCPC to CPC}
\centering
\begin{tabular}{lrrrr}
\toprule
Architecture          & cps-call & switch & condvar & spawn
\tabularnewline
\midrule
Core 2 Duo (x86-64)   &    2.45  & 1.67   & 1.13    & 2.18   \\
  Pentium~M (x86)  &  2.35 & 1.75  & 1.08  & 3.12  \\
   MIPS-32 4KEc  &  2.92 & 1.43  & 0.91  & 1.59  \\
\bottomrule
\end{tabular}
\label{tab:speed}
\end{table}

This difference of cost for cps function calls probably has an impact on the
other benchmarks, making them more difficult to interpret.  Context switches
(\emph{switch}) are around 50\,\% slower on every architecture, which is
surprisingly high since they involve almost no manipulation of environments.
Thread creation (\emph{spawn}) varies a lot across architectures: more than
3~times slower on the Pentium~M, but only 59\,\% slower on a MIPS embedded
processor.  Finally, condition variables (\emph{condvar}) are even more
surprising: not much slower on x86 and x86-64, and even 9\,\% faster on MIPS.
It is unclear which combination of factors leads eCPC to outperform CPC on this
particular benchmark only: we believe that the larger number of registers helps
to limit the number of memory accesses, but we were not able to quantify this
effect precisely.

These benchmarks of CPC primitives show that the allocation of environments
slows down eCPC in most cases, and confirms our intuition that avoiding boxing
as much as possible in favour of lambda lifting is very important in CPC.

\paragraph{Tic-tac-toe generator}

Unfortunately, benchmarking individual CPC primitives gives little information
on the performance of a whole program, because their cost might be negligible
compared to other operations.   To get a better understanding of the performance
behaviour of eCPC, we wrote a trivial but highly concurrent program with
intensive memory operations: a tic-tac-toe generator that explores the space of
grids.  It creates three threads at each step, each one receiving a copy of the
current grid, hence $3^9 = 19\,683$ threads and as many copies of the grid.

We implemented two variants of the code, to test different schemes of memory
usage.  The former is a manual scheme that allocates copies of the grids with
\texttt{malloc} before creating the associated threads, and frees each of them
in one of the ``leaf'' threads, once the grid is completed.  The latter is an
automatic scheme that declares the grids as local variables and synchronises
their deallocation with barriers;  the grids are then automatically
encapsulated, either by the boxing pass (for CPC) or in the environment (for
eCPC).

Our experiment consists in launching an increasing number of generator
\emph{tasks} simultaneously, each one generating the 19\,683 grids and threads
mentioned above.  We run up to 100 tasks simultaneously, ie.\ almost 2\,000\,000
CPC threads in total, and the slowest benchmark takes around 3~seconds to
complete on an Intel Centrino 1,87~Ghz, downclocked to 800~MHz.

Finally, we compute the mean time per tic-tac-toe task. This ratio turns out
to be independent of the number of simultaneous tasks: both CPC and eCPC scale linearly
in this benchmark.  We measured that eCPC is 20\,\% slower than CPC in the case
of manual allocation (13.2 vs. 11.0 ms per task), and 18\,\% slower in the
automatic case (31.3 vs. 26.5 ms per task).  This benchmark confirms that
environments add a significant overhead in programs performing a lot a memory
accesses, although it is not as important as in benchmarks of CPC primitives.

\paragraph{Web servers}

\begin{figure}[htbp]
\centering
\includegraphics[height=50mm]{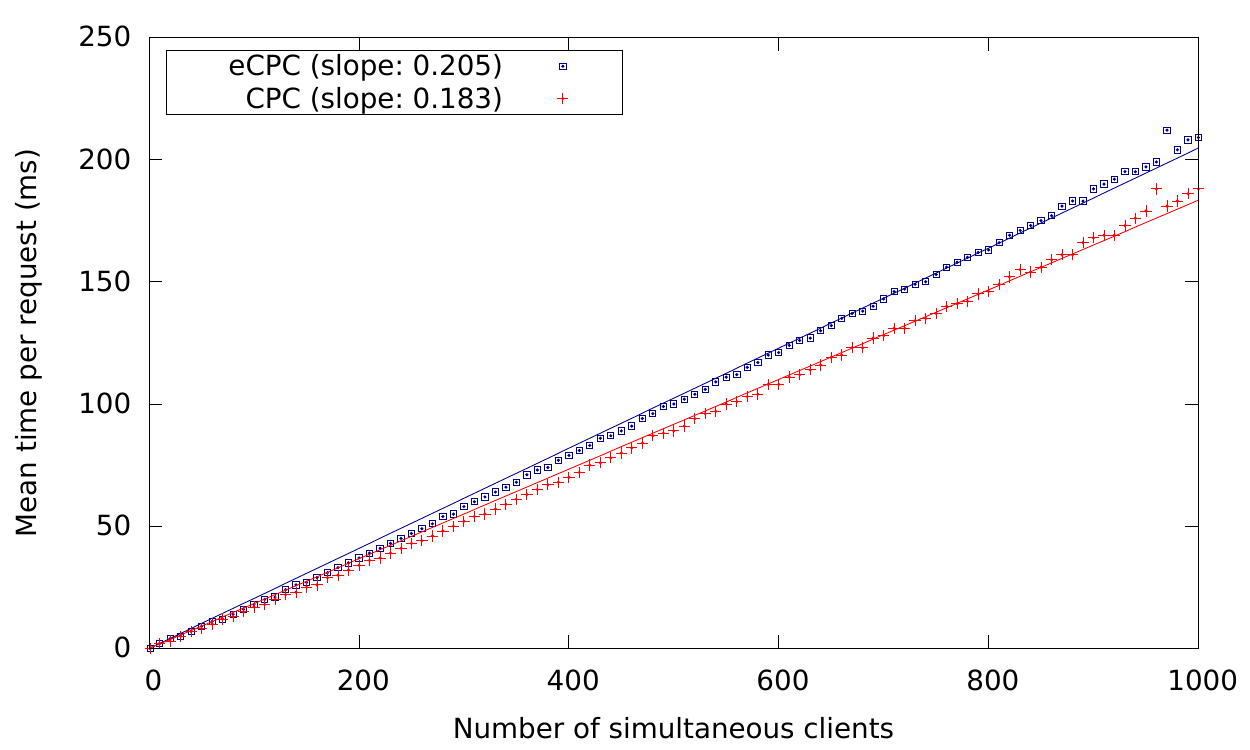}
\caption{Web server benchmark}\label{fig:ecpc-servers}
\end{figure}

To evaluate the impact of environments on more realistic programs, we reuse our
web server benchmark \cite{kerneis2009}.  We measure the mean response time of a
small web server under the load of an increasing number of simultaneous clients.
The server is deliberately kept minimal, and uses one CPC thread per client.
The results are shown in \cref{fig:ecpc-servers}.  In this benchmark, the web
server compiled with eCPC is 12\,\% slower than the server compiled with CPC.
Even on programs that spend most of their time performing network I/O, the
overhead of environments remains measurable.

\section{Conclusions}
\label{sec:conclusion}

Through the analyse of real-world programs, we have identified several typical
styles of control-flow and data-flow encodings in event-driven programs:
callbacks or state machines for the control flow, and coarse-grained or minimal
data structures for the data flow.  We have then shown how these various styles
can be generated from a common threaded description, by a set of automatic
program transformations.  Finally, we have implemented eCPC, a variant of the
CPC translator using shared environments instead of lambda lifting.  We have
found out that, although rarely used in real-world programs because it is tedious
to perform manually, lambda lifting yields better performance than environments
in most of our benchmarks.

An interesting extension of our work would be to try and reverse our program
transformations, in order to reconstruct threaded code from event-driven
programs.  This could help analysing and debugging event-driven code, or
migrating legacy, hard-to-maintain event-driven programs like Polipo towards
CPC or other cooperative threads implementations.

\subsubsection{Acknowledgments} The authors would like to thank Andy Key, who
has inspired this work, Juliusz Chroboczek for his continuous support, Peter
Sewell, Chloé Azencott and Alan Mycroft for their valuable comments.

\end{document}